\documentclass[a4paper,preprint,unsortedaddress]{revtex4-1}
\usepackage{amsmath,amssymb,amsfonts,latexsym}
\usepackage[dvips]{graphicx}
\usepackage{color}
\usepackage{indentfirst} 
\usepackage[footnotesize]{subfigure}
\usepackage{hhline}
\usepackage{booktabs}

\begin{document}

\title{Computational Efficiency and Amdahl's law for the Adaptive Resolution Simulation Technique}
\author{Christoph Junghans}
\email{junghans@lanl.gov}
\affiliation{Computer, Computational, and Statistical Sciences Division, Los Alamos National Laboratory, Los Alamos, NM 87545, USA}
\author{Animesh Agarwal}
\email{animesh@zedat.fu-berlin.de}
\affiliation{Institute for Mathematics, Arnimallee 6, D-14195, Freie Universit\"{a}t, Berlin, Germany}
\author{Luigi Delle Site}
\email{luigi.dellesite@fu-berlin.de}
\affiliation{Institute for Mathematics, Arnimallee 6, D-14195, Freie Universit\"{a}t, Berlin, Germany}
\begin{abstract}
We discuss the computational performance of the adaptive resolution technique in molecular simulation when it is compared with equivalent full coarse-grained and full atomistic simulations. We show that an estimate of its efficiency, within $10-15\%$ accuracy, is given by the Amdahl's Law adapted to the specific quantities involved in the problem. The derivation of the predictive formula is general enough that it may be applied to the general case of molecular dynamics approaches where a reduction of degrees of freedom in a multiscale fashion occurs.

\end{abstract}

\maketitle
\section{Introduction}
Adaptive Resolution Simulation (AdResS)~\cite{adress1, adress2} falls in the category of {\it ``concurrent''} multiscale methods where {\it ``concurrent''} means that a system is treated at different molecular resolutions according to the position of single molecules in space. One has one region where molecules are represented with high resolution (e.g. full atomistic) , while at the same time, in the rest of the system molecules are treated at lower resolution (e.g. coarse-grained/simple spheres).
The main characteristic of AdResS is that the exchange of particles between different regions takes place ``on-the-fly'' from one resolution to another; the technical advantage is that for some molecular systems, the region 
where the important process is taking place,  can be described in full detail (with all the explicit degrees of freedom), while 
the region far away from it, not relevant for the process of interest, can be described in less detail (coarse-grained models). A typical example is the solvation of a molecule in water; the molecule as 
well as the solvent in the first solvation shell can be treated using all the explicit degrees of freedom (atomistic resolution), 
while far away from the molecule, the solvent can be studied with satisfied accuracy using coarse-grained models. This partitioning in regions 
of different molecular resolutions has two advantages, one conceptual, that is the systematic identification of the essential degrees of freedom 
involved in a given process, and one practical, that is the drastically reduced number of degrees of freedom implies a computational gain compared to a full atomistic simulation. 
In this paper we will treat the second aspect, that is we will show the computational performance of AdResS w.r.t. the full atomistic and full coarse grained simulations. 
The paper is organized in two sections: In the first section, we provide an upper bound to the computational efficiency of AdResS based on its computational 
scaling properties. We show that the scaling properties follow the so called Amdahl's law of computational science. We will show results for the case of the AdResS 
implemented in GROMACS \cite{gromacs}, however the upper bound of efficiency is general. We generalize the formula for upper bound for AdResS systems with 
finite size of atomistic and coarse-grained regions. Thus, we have a generic formula to derive the computational gain (or, equivalently called ``speedup'') associated with AdResS simulations
compared to full-atomistic simulations. 
In the second section, we perform the numerical verification of the formula of efficiency and discuss the actual computational gain one obtains with AdResS simulations relative to full-atomistic simulations. 

\subsection{Adaptive Resolution Simulation (AdResS)} \label{adress}
\begin{figure}
   \includegraphics[width=0.75\textwidth]{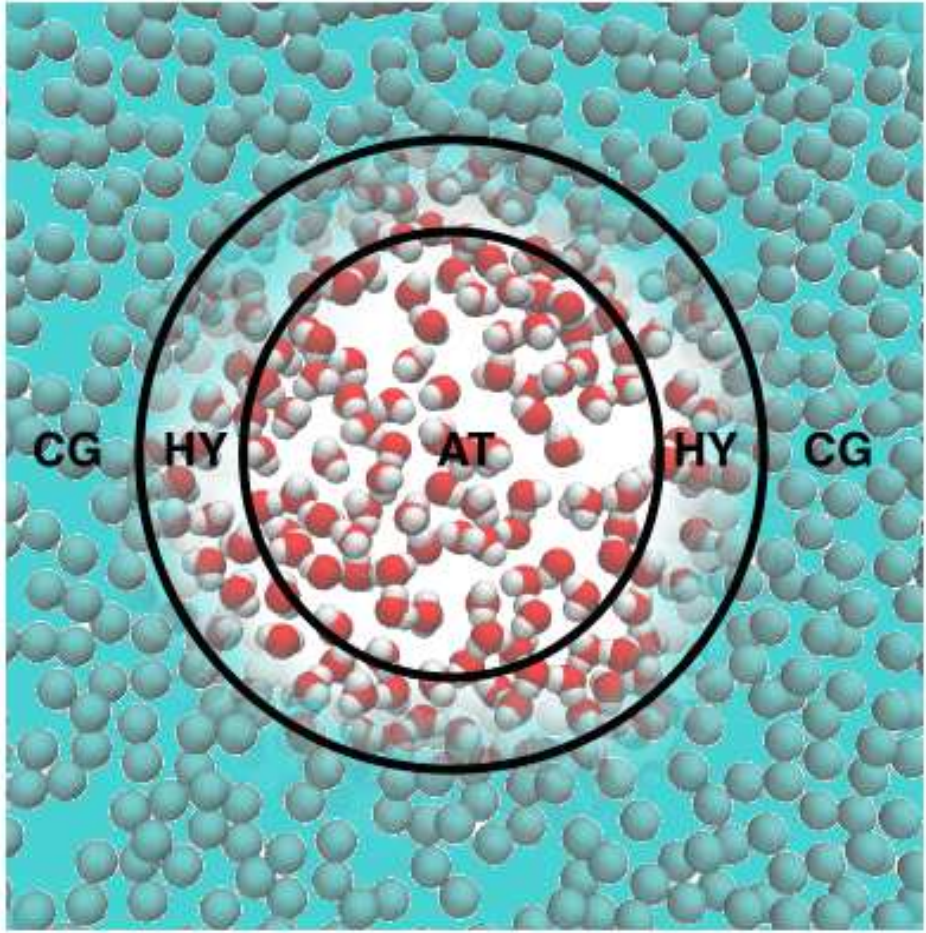}
   \caption{{Pictorial representation of the AdResS scheme with a generic molecule; CG indicates the coarse-grained region, HY the hybrid region where atomistic and coarse-grained forces are interpolated via a space-dependent, slowly varying, function $w(x)$ and AT the atomistic region (that is the region of interest).}}
   \label{fig1}
 \end{figure}

In AdResS, the simulation box is divided into three regions, one is represented by atomistic resolution, another one 
is represented by coarse-grained resolution and in between, there is a third region where the molecules 
have a mixed resolution, where both atomistic and coarse-grained 
degrees of freedom are present (see Fig.\ref{fig1}). The resolution of the molecules 
in different regions is defined by a weighting function $w(x)$. The most common weighting function that has been used in AdResS is:
\begin{equation*}
    w(x) = \begin{cases}
               1               & x < d_{AT} \\
               cos^{2}\left[\frac{\pi}{2(d_{\Delta})}(x-d_{AT})\right]   & d_{AT} < x < d_{AT}+d_{\Delta}\\
               0 & d_{AT} + d_{\Delta}< x
           \end{cases}
\end{equation*}
where, $d_{AT}$ and $d_{\Delta}$ are size of atomistic and hybrid regions respectively and x is the 
x-coordinate of the center of mass of the molecule.
The weighting function smoothly transforms from 0 to 1 in the transition region, where a coarse-grained molecule 
transforms into an atomistic molecule and 
vice versa. The molecules in atomistic and coarse-grained resolutions are coupled via an interpolation of the forces.
\begin{equation}
F_{\alpha \beta} = w(X_{\alpha})w(X_{\alpha})F_{\alpha\beta}^{atom} + [1 - w(X_{\alpha})w(X_{\alpha})]F_{\alpha\beta}^{cm}
\end{equation}
where $\alpha$ and $\beta$ indicate the two molecules, and $w(X_{\alpha})$ and $w(X_{\beta})$ are the weighting functions
characterizing these two molecules. $F_{\alpha\beta}^{atom}$ is the force in the atomistic region, which is derived 
from atomistic interactions, $F_{\alpha\beta}^{cm}$ is the force in the coarse-grained region, which is derived from 
coarse-grained potential. Thus, two molecules within the atomistic region interact via atomistic force, while 
the two molecules within the coarse-grained region interact via coarse-grained force. The rest of the molecules 
interact via spatial dependent force equation which depends on the weighting function of two molecules. The consequence is that as the molecule 
moves from the atomistic region to the coarse-grained region, the atomistic forces start diminishing, and the coarse-grained 
forces between the center of mass of the molecules start dominating. Finally, the equilibrium between the different 
regions is maintained via a global thermostat which takes care of extra energy that has to be added/removed while 
a molecule moves from the atomistic to coarse-grained region and vice versa. The above approach was shown to reproduce the 
the structural and thermodynamic properties of a wide variety of liquids within an error of $5-10$\%. For 
higher accuracy, the concept of thermodynamic force which acts on the center of mass of molecule in the hybrid region, was introduced. 
It was formulated in terms of difference of chemical potential~\cite{adress5} characterizing the atomistic region, and hybrid and coarse-grained regions. 
This approach improved the accuracy, however, it proved to be computationally expensive. The thermodynamic force was 
reformulated in terms of difference of grand potential~\cite{prl2011,jctc-han} characterizing the atomistic and coarse-grained regions and was the first step in the formulation 
Grand-Canonical like Adaptive Resolution simulations (GC-AdResS), where there is an exact Hamiltonian for the molecules contained into the atomistic region, 
while the hybrid and coarse-grained region act as reservoir of energy and particles. In Ref.~\cite{prx-han}, it was shown that the atomistic region samples the 
space in a grand-canonical fashion, and necessary conditions were derived for the probability distributions to be correct up to 
a desired order. One of the most important feature of GC-AdResS is that there is no restriction on the coarse-grained model;  
this can be just a liquid of spheres interacting via a generic potential. This was validated in Ref.~\cite{jcp-anim}, where GC-AdResS was used to calculate
the chemical potential for various liquids and mixtures and further confirmed by more elaborate and rigorous theoretical models \cite{njp}. In practice the calculation of the thermodynamic force corresponds to the equilibration of the system before the production run is initiated and it has been shown that there are rather efficient and fast ways to calculate it for a prototype system and then this same calculated force can be reused every time a system with similar characteristics is simulated ~\cite{jcp-anim,PI}. Anyway, regarding the focus of this paper, one must consider the fact that this force introduces an additional force calculation in the hybrid region and this makes the method computationally less efficient; however its application (for higher accuracy) is useful only when the hybrid region is small compared to the atomistic and to the coarse-grained region, and thus the cost of its application is computationally negligible, otherwise the standard original AdResS is more convenient.  For the reasons given above, in this paper we do not treat cases where the thermodynamic force is applied. 
\section{Amdahl's law}
Amdahl's law~\cite{amdahl} in computer science {predicts the overall speed up to a process , when only part of the computing process is improved.}
In a nutshell, if one part of a process can have a speed up due to parallelization, the overall speed up will be limited by the parts which have not gained a speed up factor from parallelization.
As an example let us consider a process in which 90\% can be sped up infinitely, but 10\% stays untouched and hence the maximum speed up is a factor 10 due to the fact that tenth of the overall run-time is remaining. In the case of AdResS the force calculation can be sped up easily, but the remaining parts cannot, we will use such similarity for a systematic estimate of the speed up (compared to a full atomistic simulation) of AdResS.

\subsection{Computational analysis}
For AdResS we will not consider absolute times, but use the time of a full atomistic simulation, $t_\text{AT}$ as a reference. In general, the running time accumulated in Molecular Dynamics (MD) simulations can be divided into two parts: 
The first part is the time needed to calculate the forces, $t_\text{F}$,
which usually goes as $N^2$ ($N$ here indicates the number of atoms) for long range interactions and $NN_\text{Neighbors}$ for short range interactions.
The second part, $t_\text{NF}$, is the time needed for performing all the other operations, such as neighbor list rebuilds, communication I/O, and even bonded interactions; this part scales with $N$.
\subsubsection{Coarse-grained simulations}
The coarse-grained simulation can be seen as the upper bound for speeding up a simulation. This is due to the fact that an AdResS simulation with vanishing atomistic region is equivalent to a coarse-grained simulation. In fact the number of atoms per molecule is reduced by clustering multiple atoms into one coarse-grained bead. 
In a coarse-grained simulations one can reach longer time scales since the intrinsic dynamics is faster (faster equilibration), larger time steps can be used due to softer interactions, and the equations of motion of a reduced number of particles need to be integrated.
Here we will not consider the first two aspects in the following calculations and limit ourselves to the speed up due to decreased number of particles (thus this estimate is a sort of ``worst case scenario'' estimate.
For a single component system, where $P$ atoms are replaced by 1 coarse-grained bead, the running time will be:
\begin{equation}
  t_\text{CG} = t_\text{F,AT}/P^2 + t_\text{NF,AT}/P
\end{equation}
This results from the fact that two beads interact with 1 instead of $P^2$ interactions, so $t_\text{F,AT}$ is reduced by a factor $P^2$ and $t_\text{NF,AT}$ is reduced by a factor $P$ due to the decreased number of particles.
For a typical simulation, for example SPC/E water, where $P=3$ and $t_\text{F}=0.75 t_\text{AT}$, one calculates:
\begin{equation}
  t_\text{CG} = 0.75 t_\text{AT}/9 +0.25 t_\text{AT} /3 = 0.167 t_\text{AT} = t_\text{AT}/6~,
\end{equation}
which means a maximum speed up factor (i.e. $\frac{t_\text{AT}}{t}$) of 6, compared to the performance of a full atomistic simulation.
This calculation is only approximating as, in real implementations, the water molecules interact with 10 (9 Coulomb and 1 Lennard-Jones) interactions instead of 9 and the coarse-grained interaction is tabulated instead of {\it hard-coded}, which implies also a slowing down factor of  $50\%$; thus a more realistic value might be 4.  {It must be noticed that the formula provided above is rather general and it is not restricted to the coarse-graining mapping of one molecule to one molecule. For example, when mapping
multiple molecules into a single coarse-grained bead the same formula can be used provided that the scaling
factor $P$ is replaced by $PK$, where $K$ is the
number of molecules mapped on a single interaction site. Furthermore, the process of bundling and unbundling of multiple molecules in AdResS implies also some computational cost; this cost, although it may not be sizable, must be anyway taken into account. In actual simulations, this procedure is still a topic of recent research and its computational optimization has not been fully developed yet \cite{bundle1,bundle2}.}
Similarly to the case of multiple mapping discussed above, in the Path Integral formulation of MD the factor $1/P^2$ has to be replaced by $1/QP^{2}$, due to the fact that atoms are treated as polymer rings and in such a case the $Q$ beads of a polymer ring of one atom interacts only with the corresponding $Q$ beads of the polymer ring of another atoms (see e.g. \cite{PI}).

\subsubsection{AdResS simulations}
The upper bound in the computational performance of AdResS  can be made more accurate by considering that the system is composed of an atomistic, a hybrid and a coarse-grained zone.
In the hybrid zone, coarse-grained as well as atomistic interactions have to be calculated.
We assume that the volume is proportional to the average number of molecules at a given (averaged) density and it follows that:
\begin{equation}
  t_\text{AdResS} = \frac{V_\text{CG}+V_\text{HY}}{V_\text{Tot}}t_\text{CG}+\frac{V_\text{AT}+V_\text{HY}}{V_\text{Tot}} t_\text{AT}.
  \label{equ:t_pieces}
\end{equation}
{Note that the above equation assumes uniform density through out the system, which is always the case in AdResS simulations.}
For $V_\text{CG}$ being much larger than $V_\text{HY}$ and $V_\text{AT}$, we again obtained $t_\text{CG}$ as an upper bound.
\subsubsection{Implementations}
In all implementations in GROMACS and Espresso, which currently exist, the atomistic representation is not actually removed in the coarse-grained region \cite{chris}.
For the moment, until a better technical solution will not be found, this is done mainly to avoid the reinsertion of the atomistic structure, which is an expensive operation from the point of view of memory. This allows also to maintain the  integration of the internal degree of freedom, which otherwise would have to be re-initialized whenever a molecule reenters the hybrid zone.
For the calculation of the speed up factor this implies that {$t_\text{CG}$ must be corrected to:}
\begin{equation}
  t^\prime_\text{CG} = t_\text{F,AT}/P^2 + t_\text{NF,AT}~,
  \label{equ:t_force_scale}
\end{equation}
which becomes the new upper bound (for the above SPC/E system, the speedup will be 3). This implies that even for very large molecules ($P \gg 1$):
\begin{equation}
  t^\prime_\text{CG} = t_\text{NF,AT}
\end{equation}
which is equivalent to Amdahl's law discussed in the previous section. For most systems $t_\text{NF,AT}$ is about $25\%$ of the total running time, so the maximum speedup will be 4.
It is important to note that this bound applies in a strict manner to the so called H-AdResS \cite{kurt}, while other version of AdResS (GC-AdResS), as underlined above may, after proper redesign of the code, avoid to keep the atomistic resolution in the coarse-grained region and hence $t_\text{NF,AT}$ can be significantly reduced. In fact, for H-AdResS, the internal degrees of freedom cannot be removed as they explicitly appear in the Hamiltonian and hence even in theory (with the best implementation possible) $t_\text{NF,AT}$ does not scale with $P$. 
Additionally $t^\prime_\text{CG}$ increases due to the use of virtual sites as coarse-grained particles (usually something of the order of $10\%$ for SPC/E water and up to $25\%$ for toluene). 
This, of course, limits the speed up even more:
\begin{equation}
  t^{\prime\prime}_\text{CG} = t_\text{NF,AT}+\frac{t_\text{NF,AT}}{t_\text{AT}}t_\text{vs}~,
  \label{equ:t_vs_correction}
\end{equation}
(assuming $t_\text{vs}$ contributes equally to $t_\text{NF,AT}$ and $t_\text{F,AT}$). For the above SPC/E system, the speedup would lowered to $4/1.1 = 3.6$.
For toluene ($P=15$), where $t_\text{F,AT}=0.74 t_\text{AT}$ and $t_\text{vs}=0.25t_\text{AT}$, the speed up will lower from 3.84 to $3.84/1.25=3.07$, 
which in is very good agreement with the numbers (obtained from simulations) reported in~\cite{toluene}.

\subsubsection{Performance Efficiency/Speed-up}
We can generalize Eqs.~\ref{equ:t_pieces},\ref{equ:t_force_scale} to AdResS systems with 
finite size of atomistic region and hybrid region and define the computational gain
one obtains when performing AdResS simulations compared to atomistic simulations.   
If we combine Eqn.~\ref{equ:t_pieces} with Eqn.~\ref{equ:t_force_scale}, then we get
\begin{equation}
	t_\text{AdResS}=\frac{V_\text{CG}+V_\text{HY}}{V_\text{Tot}}\left(t_\text{F,AT}/P^2 + t_\text{NF,AT}\right)+\frac{V_\text{AT}+V_\text{HY}}{V_\text{Tot}} t_\text{AT}
\end{equation}
and,
\begin{equation}
	t_\text{AdResS}/t_\text{AT}=v+\left(1+\frac{V_\text{HY}}{V_\text{Tot}}-v\right)\left(\%_\text{F,AT}/P^2+1-\%_\text{F,AT}\right)~,
\label{predict}
\end{equation}
where $v=(V_\text{AT}+V_\text{HY})/V_\text{Tot}$ and $\%_\text{F,AT}=t_\text{F,AT}/t_\text{AT}$. The inverse of $t_\text{AdResS}/t_\text{AT}$ is often refered to as the speedup $s$.
In conclusion the speed up factor as function of the size of the atomic region will have the functional form of $s(v)=1/(a(1+b)+v(1-a))$.
It should be noted here that the difference between a standard full-atomistic simulation and the full atomistic representation in AdResS simulation is the introduction of an additional (virtual site) in the atomistic molecule in order to couple the
coarse-grained representation to the full-atomistic one.
As a consequence a virtual size correction must be added from Eqn.~\ref{equ:t_vs_correction}; meaning scaling $1-\%_\text{F,AT}$ by $1+\%_\text{vs}$.
However, there is no way of determining {\it a priori} the cost of the virtual site correction, thus, we would need to perform the full
atomistic calculation with an extra virtual site kept at the center-of-mass of the molecule, i.e. atomistic simulations within AdResS framework. This site does not
interact with any molecules and would changed once the position and momenta of all the atoms in the molecule are
updated; therefore, in this work, we have not considered this correction.
Instead, we also report the ``actual speed up'', that is the speed up calculated by direct comparison between and AdResS simulation and a standard full atomistic simulation. This represents a further term of comparison and  a more direct practical estimate that complements the theoretical analysis of Eq.\ref{predict}. As we will see, the formula of Eq.\ref{predict} predicts always an upper bound for the ``actual speed up''.
It must be clarified that in general, there exists two approaches to study the scaling behavior of AdResS: strong scaling and weak scaling. In the former the speed up (over full-atomistic) is measured for a system of constant size varying the size of the atomistic region only. For the latter the atomistic region is keep constant and the total size of the system is varied.
The results reported in the next section concern the strong scaling approach as we believe that this approach provides a better use case for common AdResS simulation, where a full-atomistic simulation is possible but it is slow.
The weak scaling analysis and its results are instead particularly interesting for system where a very large buffer region or reservoir is needed, e.g. grand-canonical simulations \cite{prx-han,jcp-anim,njp}.
Recently strong scaling result have been reported in \cite{Kreis} looking at the force contribution only, neglecting the non-parallelized part of the code, which is the main focus of this paper.

\begin{table}[]
\begin{center}
\begin{tabular}{ccc}
\hline \hline
 System &  $\%_\text{F,AT}$ \\
\hline
TIP5P water & 63 \\
butanol & 58  \\
Hexane & 46  \\
DMSO & 55 \\
TBA-DMSO mixture & 57 \\
\hline \hline
\end{tabular}
\caption{Estimate of \%time spent in force calculation in a full atomistic simulation of various systems studied in this work. It must be noticed that this information is given by the code in terms of generic time required for the calculation of forces. In reality given the complexity of the computational architecture and the entanglement between overlapping operations the real time for force calculations may be smaller. In fact the factor of $10-15 \%$ of disagreement between the theoretical estimate and the numerical results carries in part the uncertainty of \%time spent in force calculation. In any case, we assume to be in an ideal condition where the time printed by the code is indeed the time of force calculations only. Such an assumption implies that our theoretical estimate can be viewed always as an indication (within  $10-15 \%$) of the best performance (upper bound) of the real computational calculation.}
\label{force}
\end{center}
\end{table}

\begin{table}[htb]
\centering
\begin{minipage}{0.48\textwidth}
\sffamily
\begin{tabular}{l*{5}{c}}
\toprule
\hline \hline
 \bfseries $AT$ (nm) & \bfseries v & \bfseries Equation & \bfseries Simulation \\
\midrule
\hline \hline
0.0 & 0.0 & 2.53 & 2.94\\
0.5 & 0.0136 & 2.45 & 2.69  \\
0.9 & 0.0284 & 2.37 & 2.47\\
1.3 & 0.0511 & 2.26 & 2.17 \\
1.7 & 0.0835 & 2.13 & 1.88\\
2.1 & 0.127 &  1.98 & 1.60\\ \bottomrule
\hline \hline
\end{tabular}
\caption{Comparison of speed up for TIP5P water using IBI potential  \cite{ibi} with pressure correction in the coarse-grained region, calculated using Eq. 9 and from AdResS simulation. ``AT'' refers to the size of the atomistic region, $v=(V_\text{AT}+V_\text{HY})/V_\text{Tot}$, the third column reports the quantity $\frac{t_\text{AT}}{t_\text{AdResS}}$ from the formula and the fourth column reports the quantity $\frac{t_\text{AT}}{t_\text{AdResS}}$ from simulation; the same convention is used in all the following tables.}
\label{tip5p}
\end{minipage}%
\hfill
\begin{minipage}{0.48\textwidth}
\centering
\sffamily
\begin{tabular}{l*{3}{c}}
\toprule
\hline \hline
 \bfseries $AT$ (nm) & \bfseries Actual Speedup \\
\midrule
\hline \hline
0.0 & 2.45\\
0.5 & 2.24  \\
0.9 & 2.05\\
1.3 & 1.81 \\
1.7 & 1.56\\
2.1 & 1.33\\ \bottomrule
\hline \hline
\end{tabular}
\caption{{Actual speed up $\frac{t_\text{AT}}{t_\text{AdResS}}$ for TIP5P water using IBI potential with pressure correction in the coarse-grained region. ``AT'' refers to the size of the atomistic region.}}
\label{tip5p1}
\end{minipage}
\end{table}

\begin{table}[htb]
\centering
\begin{minipage}{0.48\textwidth}
\sffamily
\begin{tabular}{l*{5}{c}}
\toprule
\hline \hline
 \bfseries $AT$ (nm) & \bfseries v & \bfseries Equation & \bfseries Simulation \\
\midrule
\hline \hline
0.0 & 0.0 & 2.28 & 2.07\\
0.5 & 0.0127 & 2.22 & 1.98\\
1.0 & 0.0265 & 2.16 & 1.92\\
1.3 & 0.0476 & 2.08 & 1.82\\
1.7 & 0.0779 & 1.97 & 1.71\\
2.1 & 0.118 & 1.85 & 1.56\\ \bottomrule
\hline \hline
\end{tabular}
\caption{Comparison of speed up for liquid butanol using IBI potential with pressure correction in the coarse-grained region, calculated using Eq. 9 and from AdResS simulation.}
\label{butanol}
\end{minipage}%
\hfill
\begin{minipage}{0.48\textwidth}
\centering
\sffamily
\begin{tabular}{l*{3}{c}}
\toprule
\hline \hline
 \bfseries $AT$ (nm) & \bfseries Actual Speedup \\
\midrule
\hline \hline
0.0 & 2.02\\
0.5 & 1.93  \\
0.9 & 1.87\\
1.3 & 1.77 \\
1.7 & 1.66\\
2.1 & 1.52\\ \bottomrule
\hline \hline
\end{tabular}
\caption{{Actual speed up $\frac{t_\text{AT}}{t_\text{AdResS}}$ for liquid butanol using IBI potential with pressure correction in the coarse-grained region. ``AT'' refers to the size of the atomistic region.}}
\label{butanol1}
\end{minipage}
\end{table}

\begin{table}[htb]
\centering
\begin{minipage}{0.48\textwidth}
\sffamily
\begin{tabular}{l*{5}{c}}
\hline \hline
 \bfseries $AT$ (nm) & \bfseries v & \bfseries Equation & \bfseries Simulation \\
\midrule
\hline \hline
0.0 & 0.0 &  2.06 & 1.88 \\
0.5 & 0.0167 & 2.00 & 1.78\\
0.9 & 0.0348 & 1.93 & 1.70\\
1.3 & 0.0627 & 1.85 & 1.60\\
1.7 & 0.102 & 1.74 & 1.46\\
2.1 & 0.156 & 1.62 & 1.33\\ \bottomrule
\hline \hline
\end{tabular}
\caption{Comparison of speed up for liquid DMSO using IBI potential with pressure correction in the coarse-grained region, calculated using Eq. 9 and from AdResS simulation.}
\label{dmso}
\end{minipage}%
\hfill
\begin{minipage}{0.48\textwidth}
\centering
\sffamily
\begin{tabular}{l*{3}{c}}
\toprule
\hline \hline
 \bfseries $AT$ (nm) & \bfseries Actual Speedup \\
\midrule
\hline \hline
0.0 & 1.66\\
0.5 & 1.57  \\
0.9 & 1.50\\
1.3 & 1.41 \\
1.7 & 1.28\\
2.1 & 1.17\\ \bottomrule
\hline \hline
\end{tabular}
\caption{{Actual speed up $\frac{t_\text{AT}}{t_\text{AdResS}}$ for liquid DMSO using IBI potential with pressure correction in the coarse-grained region. ``AT'' refers to the size of the atomistic region.}}
\label{dmso1}
\end{minipage}
\end{table}

\begin{table}[htb]
\centering
\begin{minipage}{0.48\textwidth}
\sffamily
\begin{tabular}{l*{5}{c}}
\hline \hline
 \bfseries $AT$ (nm) & \bfseries v & \bfseries Equation & \bfseries Simulation \\
\midrule
\hline \hline
0.0 & 0.0 &  1.8 & 2.06\\
0.5 & 0.0113 & 1.76 & 2.00\\
0.9 & 0.0236 & 1.73 & 1.93\\
1.3 & 0.0425 & 1.68 & 1.85 \\
1.7 & 0.0694 & 1.63 & 1.73\\
2.1 & 0.105 & 1.56 & 1.62\\ \bottomrule
\hline \hline
\end{tabular}
\caption{Comparison of speed up for liquid hexane using IBI potential with pressure correction in the coarse-grained region, calculated using Eq. 9 and from AdResS simulation.}
\label{hexane}
\end{minipage}%
\hfill
\begin{minipage}{0.48\textwidth}
\centering
\sffamily
\begin{tabular}{l*{3}{c}}
\toprule
\hline \hline
 \bfseries $AT$ (nm) & \bfseries Actual Speedup \\
\midrule
\hline \hline
0.0 & 1.45\\
0.5 & 1.41  \\
0.9 & 1.36\\
1.3 & 1.30 \\
1.7 & 1.22\\
2.1 & 1.14\\ \bottomrule
\hline \hline
\end{tabular}
\caption{{Actual speed up $\frac{t_\text{AT}}{t_\text{AdResS}}$ for liquid hexane using IBI potential with pressure correction in the coarse-grained region. ``AT'' refers to the size of the atomistic region.}}
\label{hexane1}
\end{minipage}
\end{table}

\begin{table}[htb]
\centering
\begin{minipage}{0.48\textwidth}
\sffamily
\begin{tabular}{l*{5}{c}}
\hline \hline
 \bfseries $AT$ (nm) & \bfseries v & \bfseries Equation & \bfseries Simulation \\
\midrule
\hline \hline
0.0 & 0.0 & 2.17 & 2.16\\
0.9 & 0.0166 & 2.10 & 2.01\\
1.3 & 0.0298 & 2.05 & 1.96 \\
1.7 & 0.0488 & 1.98 & 1.88\\
2.1 & 0.0744 & 1.91 & 1.76\\ 
2.5 & 0.107 & 1.82 & 1.65 \\ \bottomrule
\hline \hline
\end{tabular} 
\caption{Comparison of speed up for TBA/DMSO mixture using IBI potential with pressure correction in the coarse-grained region, calculated using Eq. 9 and from AdResS simulation.}
\label{mixture}
\end{minipage}%
\hfill
\begin{minipage}{0.48\textwidth}
\centering
\sffamily
\begin{tabular}{l*{3}{c}}
\toprule
\hline \hline
 \bfseries $AT$ (nm) & \bfseries Actual Speedup \\
\midrule
\hline \hline
0.0 & 1.56 \\
0.9 & 1.45  \\
1.3 & 1.41 \\
1.7 & 1.35 \\
2.1 & 1.27 \\
2.5 & 1.19 \\ \bottomrule
\hline \hline
\end{tabular}
\caption{{Actual speed up $\frac{t_\text{AT}}{t_\text{AdResS}}$ for TBA/DMSO mixture using IBI potential with pressure correction in the coarse-grained region. ``AT'' refers to the size of the atomistic region.}}
\label{mixture1}
\end{minipage}
\end{table}

\subsubsection{Case of Mixtures}
{In the case of mixture $t_\text{F,AT}$ has to be split into the contributions from the different species of molecules.
If we consider a mixture with two types of molecules $A$ and $B$, then $t_\text{F,AT}$ can be written in terms 
of contribution of different types of interactions.
\begin{equation}
	\frac{t_\text{F,AT}}{t_\text{AT}}=\%_\text{F,AT,AA}+\%_\text{F,AT,AB}+\%_\text{F,AT,BB}
\end{equation}
where $t_\text{AT}$ is the time for the full-atomistic simulation, $\%_\text{F,AT,AA}$ is the 
$\%$ time spent in force calculation when only $A-A$ interactions are switched on and rest of the 
interactions are switched off. Similarly, one can define $\%_\text{F,AT,AB}$ and $\%_\text{F,AT,BB}$.
Since there is no direct procedure to calculate the individual percentages, we first calculate $\%_\text{F,AT, mix}$, i.e., 
$\%$ time spent in force calculation in $A-B$ mixture, which can be obtained directly from GROMACS output. We 
then calculate $\%_\text{F,AT,AA}$, $\%_\text{F,AT,AB}$ and $\%_\text{F,AT,BB}$ by considering the number of interactions of
specific type in the system. Suppose there are $X$ number of interactions of type $A-A$, $Y$ number of interactions of type $A-B$ and
$Z$ number of interactions of type $B-B$, then 
\begin{equation}
\%_\text{F,AT,AA} = \frac{X}{X+Y+Z} \%_\text{F,AT, mix}
\end{equation}
\begin{equation}
\%_\text{F,AT,AB} = \frac{Y}{X+Y+Z} \%_\text{F,AT, mix}
\end{equation}
\begin{equation}
\%_\text{F,AT,BB} = \frac{Z}{X+Y+Z} \%_\text{F,AT, mix}
\end{equation}
To obtain the coarse-grained time, these contributions need to be scaled by the number of atoms which are coarse-grained:
\begin{equation}
	\frac{t_\text{F,CG}}{t_\text{AT}}=(\frac{\%_\text{F,AT,AA}}{P_\text{A}^2}+\frac{\%_\text{F,AT,AB}}{P_\text{A}P_\text{B}}+\frac{\%_\text{F,AT,BB}}{P_\text{B}^2})
\end{equation}
where $P_\text{A}$ is the number of atoms coarse-grained in molecule $A$ and $P_\text{B}$ is the number of atoms coarse-grained in molecule $B$.
}
\subsection{Numerical Results}
We compare the computational gain obtained using Eq.~\ref{predict} with the actual gain
obtained from AdResS simulations compared to the full-atomistic simulation. We treated 
five different systems in this work: liquid water, hexane, butanol, DMSO and TBA/DMSO mixture.  
The technical details of these systems as well as simulations details are reported in 
Appendix. All the simulations are performed on a single processor Intel Xeon CPU E31245 machine,  
since the load balancing in AdResS and atomistic simulations works differently.
Table~\ref{force} shows the percentage of time that is spent in force calculation ($\%_\text{F,AT}$) 
for the above mentioned systems and Table~\ref{tip5p}, Table~\ref{butanol},  Table~\ref{dmso},  Table~\ref{hexane} and Table~\ref{mixture} show the speed up 
obtained with AdResS simulations of these systems relative to {atomistic systems within AdResS framework (i.e. with an additional virtual site).
Table~\ref{tip5p1}, Table~\ref{butanol1},  Table~\ref{dmso1},  Table~\ref{hexane1} and Table~\ref{mixture1} show the actual speedup obtained with AdResS simulations 
relative to full-atomistic simulations. Since there is no additional cost of virtual site involved, the actual speedup is lower.}
We have used TIP5P model for water; the reason for this choice was made in order to make a consistent comparison between the performance of AdResS and that of a full atomistic simulation. In fact the full-atomistic simulations for standard water models such
as SPC, SPC/E in GROMACS are highly optimized, while AdResS in GROMACS, for the moment, is not equivalently optimized for such models. It must 
be noted here that we have only varied the size of the atomistic region, while keeping the 
thickness of the hybrid region fixed. While the overall trend of the numerical results follows the theoretical prediction of Amdahl's law, it can be seen that the maximum discrepancy between the predicted 
speedup and the speed up obtained from the simulations is 15 \% (but in some cases, e.g. hexane, the 
maximum discrepancy is less than 10 \%). One of the reasons of such a discrepancy is the use of coarse-grained  potential (obtained with the Iterative Boltzmann Inversion, IBI,technique with
pressure correction) in the coarse-grained region. In fact, Eq.~\ref{predict} assumes that the (average) density is uniform across 
the atomistic, hybrid and coarse-grained region, however, it has been reported (and discussed) that the 
density in the hybrid region is not perfectly uniform and presents a systematic discrepancy of 5\% w.r.t. the reference density of the full-atomistic simulation; this leads to the fact that the particle density in the
atomistic and coarse-grained regions is not the same as that of target (i.e. that of the full-atomistic simulation of reference). Although the discrepancy regarding the density is small, 
and it does not affect the structural and thermodynamic properties, this difference could affect the ``ideal'' performance (of Amdahl's Law) that one may expect from an AdResS simulation. As underlined before, from the technical point of view the introduction of the thermodynamic force removes the problem of non homogeneous density, but also implies an additional force calculation in the hybrid region, this extra calculation is not massive and in principle can be neglected, for relatively small hybrid regions, in the estimate of the performance, however we prefer to remain within the {\it ``worst case scenario''} of non uniform density in order to estimate the maximum discrepancy of the numerical results compared to the theoretical prediction of Amdahl's formula.
Additionally it is important to remember that the IBI potential are tabulated potentials, while the full-atomistic simulation uses Lennard-Jones type potential and there is a certain level of penalty (~5\%) for using those. 
{On the other hand for some of the atomistic systems (e.g. TIP5P), Coulomb interactions need to be calculated as well; this aspect have not been explicitly considered but it is assumed that the interactions between two atoms, respectively beads in CG simulation, lead to the same order of calculation time.}
{Finally, the additional virtual site adds additional overhead to the AdResS simulation in comparison to the full atomistic simulation.} Thus, we can use the formula of Amdahl's law to estimate the upper bound for the gain in computational efficiency of AdResS compared to an equivalent full atomistic simulation but must keep in mind an uncertainty of about $10-15\%$.
However, we must also keep in mind that the estimate formula suggested here regards the {``worst scenario''} case of computational implementation of AdResS, i.e. the only part affected is in the calculation of the forces. In practice, there are several aspects where the computational architecture can be improved and further additional efficiency improvements can be added to the one discussed in this work.

\section{Conclusions}
We have developed a model to predict the upper bound of the speedup which can be achieved in an adaptive resolution simulations and look at the particular case of AdResS.
First, it is obvious that a full coarse-grained simulation is the maximum speedup possible.
Second, assuming that all the code implementations of the AdResS method are not ideal, that is, they only optimize the force calculation, the upper bound for the speed up is controlled within a limit imposed by Amdahl's law and hence even if the force calculation would not have a cost, which is nearly true for large molecules, a maximum speedup inversely proportional to the non-force calculation part of the full-atomistic simulation is imposed. We have tested this hypothesis on multiple systems and confirmed that performance model within 15\% accuracy, which is reasonable considering the broad range of systems and the simplicity of the performance model (this latter does not incorporate any correction terms related to the inhomogeneous density, and other different interaction types used in the full-atomistic simulation). Finally, from the various calculations can be seen that for reasonable sizes of the atomistic region (at fixed coarse-grained region) one can have a gain factor of about 1.5-2.0. While for small systems this is not relevant, for large systems it may represent a sizable gain, for example in the path integral based AdResS \cite{jcp-anim,cmd} such a gain allows for treating systems otherwise at the limit of the computational capability of standard resources. In this perspective, a formula as the one proposed here allows researchers to estimate the gain they would have if they treat the system of interest. To decide whether or not to use AdResS, if the aim is to save computational time, can be made practical by our proposal.

\section*{Acknowledgments}
 This work was supported by the Deutsche Forschungsgemeinschaft (DFG) through grant CRC 1114 (project C01) and by the European Community through project E-CAM awarded to L.D.S.; calculations were performed using the computational resources of the North-German Supercomputing Alliance (HLRN), project {\bf bec00127}.
\appendix
\section{Technical details}
\label{app:i}
All the simulations are performed with Gromacs-4.6.7. A spherical atomistic and transition region 
are used in all the AdresS simulations. The time step is used in all the simulations is 2 fs. 
For each system, the size of the atomistic region is varied while the size of the transition region (0.95 nm) is kept fixed. We use 
a Langevin thermostat with inverse friction coefficient 0.1 ps in all the simulations, and the 
temperature is maintained at 300 K. The reaction-field method 
is used to treat the electrostatic interactions, while the cut-off method is used to treat the 
Vander-Walls interactions. We use TIP5P model for water molecule and the parameters are taken from
OPLSAA force field. The parameters for Liquid Hexane,butanol and DMSO are taken from 
GROMOS53A6 parameter set. For all the systems, a cut-off radius of
0.9 nm was used for van der Waals and Coulomb interactions. To obtain the initial trajectory to be used 
in AdResS simulations, we performed an equilibration run under NPT conditions for 1 ns.   
To obtain the coarse-grained potential, we used Iterative Boltzmann Inversion in conjunction 
with pressure correction, which ensures that the coarse-grained region has the same 
pressure as the atomistic region. Table~\ref{tec} lists the specific 
details of the systems studied in this work. 
\begin{table}[]
\begin{center}
\begin{tabular}{ccc}
\hline \hline
 System & $N_{atom}$ & System size ($\textrm{nm}^{3}$)  \\
\hline
TIP5P water & 30000 & 9.77$\times$9.77$\times$9.77 \\
butanol & 6530 & 10.01$\times$10.01$\times$10.01 \\
Hexane & 4700 & 10.39$\times$10.39$\times$10.39 \\
DMSO & 6020 & 9.12$\times$9.12$\times$9.12 \\
TBA/DMSO & 2500/10000 & 11.69$\times$11.69$\times$11.69 \\
\hline \hline
\end{tabular}
\caption{Technical details of the various systems studied in this work}
\label{tec}
\end{center}
\end{table}

\end{document}